\documentclass[prd,superscriptaddress,nofootinbib,showpacs,preprint]{revtex4}

\usepackage{graphicx}
\usepackage{color}
\usepackage{amsmath}
\usepackage{amsfonts}
\usepackage{amssymb}
\usepackage[normalem]{ulem}
\usepackage[colorlinks=true,linkcolor=blue,citecolor=blue]{hyperref}
\usepackage{subfigure}
	
\begin{document}
		
\title{Viscous Self Interacting Dark Matter Cosmology \\ For Small Redshift} 
\author{Abhishek Atreya }
\email{atreya.abhi@gmail.com}
\affiliation{Center For Astroparticle Physics and Space
    Sciences, Bose Institute, Kolkata, 700009, India}
\author{Jitesh R. Bhatt }
\email{jeet@prl.res.in}
\affiliation{Theoretical Physics Division, Physical Research Laboratory,
	Navrangpura, Ahmedabad, 380009, India}	
\author{Arvind Kumar Mishra }
\email{arvind@prl.res.in}
\affiliation{Theoretical Physics Division, Physical Research Laboratory,
	Navrangpura, Ahmedabad, 380009, India}
\affiliation{Indian Institute of Technology, Gandhinagar, 382424, India }

\begin{abstract}
The viscosity of dark matter in cosmological models may cause an accelerated expansion and when this effect is sufficiently large, it can explain the dark energy. In this work,  attributing the origin of viscosity to self-interaction of dark matter, we study the viscous cosmology at small redshift $(0\leq z\leq2.5)$. Assuming the cluster scale to be virialized and by modeling a power law behavior of velocity gradients,  we calculate the Hubble expansion rate, $H(z)$ and the deceleration parameter, $q(z)$. We then perform a $\chi^{2}$ analysis to estimate the best fit model parameters. By using the best fit values, we explain the cosmic chronometer and type Ia supernova data. We conclude that if the dissipative effects become prominent only at the late time of cosmic evolution and are smaller at higher redshift, we can explain the observational data without requiring any dark energy component. Our analysis is independent of any specific model of self interacting dark matter.

\end{abstract}

		\maketitle
		\pagebreak

\section{Introduction}

The observations of large redshift
supernovae provide a
compelling evidence that the Universe has gone in an
accelerating phase lately
\cite{Riess:1998cb,Perlmutter:1998np}.
To explain the observations, a hypothetical new form of energy, called the
dark energy (DE), is required which contributes $\sim 70$\% of the
total energy budget of the Universe. Rest $\sim30$\% of the energy density comes
 from  the baryonic matter and an unknown form of matter,
called dark matter (DM). Several other independent
  observations such as Baryon Acoustic Oscillations (BAO)
  \cite{Eisenstein:2005su}, Cosmic Microwave Background Radiation 
  (CMBR) \cite{Komatsu:2010fb}, Large Scale Structures (LSS)
  \cite{Tegmark:2006az} etc. provide concrete support to the above picture.

Many theoretical attempts have been made to explain dark
energy either by modifying the gravity sector or the energy momentum content of the Einstein's field equation, see
review \cite{Copeland:2006wr,Yoo:2012ug}. 
The cosmological constant $\Lambda$, is
the most successful model  to explain dark energy but
there is no theory to explain its physical origin
and its present value, see Ref. \cite{Weinberg:1988cp}.

It is a possibility that the dark energy
is not a separate entity but
a manifestation of some
intrinsic properties of dark matter. It has been argued that if cosmic fluid
is not perfect but viscous, then its effective pressure, $P+\Pi_{B}$  may 
turn negative and change the solution of Einstein's equation,
thereby affecting the cosmic
evolution of the Universe, leading to an
inflation like behaviour
\cite{Padmanabhan:1987dg, Gron:1990ew, Cheng:1991uu, Zimdahl:1996ka}.
Similar arguments have been made in the context of 
currently observed cosmic acceleration
\cite{Fabris:2005ts, Avelino:2008ph, Mohan:2017poq, Cruz:2018yrr}. 
It have been also shown that the bulk viscosity can unify the dark sectors.
i.e. dark matter and dark energy. The source of bulk viscosity
  has been attributed to neutrinos \cite{Das:2008mj}, an exotic scalar field 
\cite{Gagnon:2011id} or the decay of cold dark matter into relativistic
particles \cite{Mathews:2008hk}. However these model are highly constrained
by CMB and LSS observations, as  large bulk viscosity reduces
  the gravitational potential, which in turn affect the structure formation
\cite{Li:2009mf,Barbosa:2015ndx, Piattella:2011bs,Velten:2011bg}. 
For recent discussions on the role of cosmic viscosity, we refer to 
\cite{Anand:2017wsj,Anand:2017ktp, Goswami:2016tsu, Cai:2017buj,Lu:2018smr,Brevik:2017msy}.

However it has been argued in Ref. \cite{Floerchinger:2014jsa} that at late
times, both bulk and shear viscosity of cosmic fluid might play an important
role in cosmic expansion  and contribute to the observed accelerated expansion
of the Universe without any dark energy. In our previous work
\cite{Atreya:2017pny}, we argued that dark matter self-interactions can  produce sufficient dissipation that can explain the present cosmic acceleration. The motivation for SIDM comes from
the small scale astrophysical observations which demand some self-interactions
between the dark matter particles \cite{Spergel:1999mh,Tulin:2017ara}. Due to
self-interactions it is natural to expect some non-zero viscosity in the dark
matter sector.  Also SIDM does not conflict with the LSS observation, as
it behaves like an interacting matter at small scales and a non-interacting
matter at large scales. 

We derive the expressions for viscous coefficients (shear and bulk viscosity) of SIDM using the kinetic theory formalism and
estimate the bulk and shear viscosity due to SIDM. To estimate the viscosity, we
use the value of $\langle\sigma v\rangle/m$ as obtained in Ref.
\cite{Kaplinghat:2015aga} by utilizing the astrophysical data from dwarf
galaxies, Low Surface Brightness (LSB) galaxies and clusters. Further we
estimate the mean free path of SIDM $\lambda_{\mathrm{SIDM}}\sim 1$ Mpc, which is
order
of cluster scale and argue that the cluster scale is smallest scale where
the hydrodynamics is valid. We concluded that the viscous effects of the
SIDM are sufficient enough to account for the present observed accelerated
expansion of the Universe without any extra dark energy component
\cite{Atreya:2017pny}.

 In this work we extend our previous study to look at the
  late time dynamical evolution of universe within the framework of viscous
  SIDM cosmology without any dark energy component
  \cite{Atreya:2017pny}. An
  important assumption in our analysis is that in recent times the clusters
  size dark matter halos have virialized and thus these are the relevant scales
  for estimating viscosity. At larger scales the velocity perturbations evolve
  and the viscous effects thus manifest themselves at those scales by
  contributing to energy dissipation. To calculate the dissipation
  term, we consider a few simplified assumptions.
  Assuming a power law parameterization 
  of average velocity gradient on the redshift and with our estimates of
  viscosity of SIDM
  from \cite{Atreya:2017pny}, we set-up the
equations for the
  Hubble parameter $H(z)$ and deceleration parameter $q(z)$.
  
  To extract the values of power law exponent and
  length scale,
  we use the $ \chi^{2} $ analysis with the cosmic chronometer data and find that the best fit values
also explain the supernova data.
 The best fit values extracted from the fit dictate that
    dissipation was smaller at earlier times which is consistent with the
    expectation that the gradients become prominent only at late times thus
    affecting the cosmic evolution. We
  also extract the epoch of deceleration-acceleration transition
  and find that the value of deceleration parameter approaches $q\sim 0.5$, for
  the best fit values, in the matter dominated era as expected.

The arrangement of our work is as follows: 
In section \ref{sec:first}, using the kinetic theory formalism we calculate
the viscosity of SIDM and discuss its effect on the cosmic evolution. In
section \ref{sec:dcal}, following some simplifying assumptions, we approximate
the form of dissipation term $D$. By using the from of $D$, we then set up the
coupled differential equation for the Hubble rate and deceleration parameter.
In
  section \ref{sec:chi} we perform the $\chi^{2}$ analysis with the cosmic
chronometer data and estimate the best fit model parameter of viscous SIDM
model. In section \ref{sec:results}, we present our results. The best fit
values successfully explain supernova data. We also discuss the evolution of
deceleration parameter and dissipation due to viscous effects. In the last
section \ref{sec:danc}, we summarize and conclude our work.
 
Throughout the paper, we 
adopt the convention for
denoting differentiation 
 $\dot{A} \equiv \frac{dA}{d\tau} $  and 
$A'\equiv \frac{dA}{dz} $.

\section{Viscosity of SIDM and Cosmic Acceleration}\label{sec:first}
In this section, we estimate the viscosity of SIDM due to self
  interactions, using kinetic theory \cite{Atreya:2017pny}. We then use it to estimate the
  corrections to the Einstein equations. 
\subsection{Viscosity of SIDM}
\label{sec:visc}
The starting point of kinetic theory is Boltzmann's equation
\begin{equation}
\label{eq:be}
\frac{\partial f_{p}}{\partial t} + v_{p}^{i}\frac{\partial f_{p}}{\partial x^{i}} = I\lbrace f_{p}\rbrace,
\end{equation}
where $\mathbf{v}_{p}$ is single particle velocity, $f_{p}$ is the distribution
function and $I\lbrace f_{p}\rbrace$ is collisional term. Within the
``relaxation time formalism'', we can approximate the change in the
distribution function due to collisions as
\begin{equation}
\label{eq:delf}
\delta f_{p} = - \tau \left(\frac{\partial f_{p}^{0}}{\partial t} +
v_{p}^{i}\frac{\partial f_{p}^{0}}{\partial x^{i}} \right).
\end{equation}  
where $\tau$ is the relaxation time. It was argued in Ref. \cite{Atreya:2017pny}
that this is a valid approximation for dark matter halos. Since $\delta f_{p}$
is the deviation from the equilibrium distribution, the dissipative part of
total energy momentum tensor, $T^{\mu\nu}=T^{\mu\nu}_{\mathrm{Ideal}} +
T^{\mu\nu}_{\mathrm{Diss}}$, is given by
\begin{equation}
\label{eq:emt}
T^{ij}_{\mathrm{Diss}} = \int\frac{d^{3}p}{(2\pi)^{3}}~v^{i}p^{j}\delta f_{p}~~.
\end{equation}
Comparing with the dissipative component of $T^{\mu\nu}$ in viscous
hydrodynamics one finds \cite{Gavin:1985ph,Kadam:2015xsa} 
\begin{equation}
\label{eq:visc}
\eta = \frac{1}{15T}\int \frac{d^{3}p}{(2\pi)^{3}}~\tau~\frac{p^{4}}{E_{p}^{2}}
~\frac{\partial f_{p}^{0}}{\partial E_{p}}~~,
\end{equation}  
\begin{equation}
\label{eq:blk}
\mathrm{and\quad}\zeta = \frac{1}{T}\int\frac{d^3p}{(2\pi)^3}\tau\left[E_{p}C_{n}^2 - \frac{p^2}{3E_{p}}\right]^2 f_{p}^{0}~~.
\end{equation}
where $C_{n} = \frac{\partial P}{\partial \epsilon}\lvert_{n}$ is the speed of
sound at constant number density.

The exact expression for $\tau$ will depend on the specific
  model of SIDM. To keep our analysis model independent, we approximate relaxation time $\tau$, by its 
  thermal average
\begin{equation}
\label{eq:avt}
\tilde{\tau}^{-1} = n \langle\sigma v\rangle,
\end{equation}
where $n, ~\langle\sigma v\rangle$ are the average number density and
velocity weighted cross-section average respectively. Using the
non-relativistic Maxwell-Boltzmann distribution in fluid rest frame, with eq.
(\ref{eq:avt}), in the non-relativistic limit of eq. (\ref{eq:visc}) and
(\ref{eq:blk}) we get \cite{Atreya:2017pny}
\begin{equation}
\label{eq:visc2}
\eta =   \frac{1.18m\langle v \rangle^{2}}{3\langle\sigma v\rangle}.
\end{equation}
\begin{equation}
\label{eq:blk2}
\zeta =\frac{5.9 \ m\langle v\rangle^{2}}{9\langle\sigma v\rangle}.
\end{equation}
We have used equipartition of energy to relate root mean square velocity with
the temperature $T$ in deriving eq. (\ref{eq:visc2}) and (\ref{eq:blk2}). For
more details on the calculations, see Ref. \cite{Atreya:2017pny}.
\subsection{Einstein's Equation With Viscosity}
\label{sec:eeqns}
In this section, we investigate the effects of viscosity of
  SIDM on the solution of Einstein's equation
  \cite{Floerchinger:2014jsa,Atreya:2017pny}. We consider that our Universe  
  consists of the viscous dark matter with no extra dark
  energy component.
  
The starting point is the total energy momentum tensor
  $T^{\mu \nu}  $ of viscous fluid. 
In the Landau frame, one can write the energy momentum tensor for
  the viscous dark matter in the first order gradient expansion as
 \begin{equation}
 T^{\mu \nu} = \epsilon u^{\mu} u^{\nu} + (P + \Pi_{B}) \Delta^{\mu \nu} +\Pi^{\mu \nu},
 \end{equation}
where $ \Delta^{\mu \nu} = u^{\mu} u^{\nu} + g^{\mu \nu}$
 is the projection operator and  $\Pi_{B}$, $\Pi^{\mu \nu}$ represent
 bulk stress and shear stress 
tensor respectively, defined as
 \begin{equation}
 \Pi_{B} = - \zeta \nabla_{\mu} u^{\mu}
 \end{equation}
 \begin{equation}
  \mathrm{and}\quad \Pi^{\mu \nu}= - \eta\left[ \Delta^{\mu\alpha}\Delta^{\nu\beta} +
     \Delta^{\mu\beta}\Delta^{\nu\alpha} - \frac{2}{3}\Delta^{\mu\nu}
     \Delta^{\alpha\beta}\right] \nabla_{\alpha}u_{\beta}~~,
 \end{equation}
where $\eta$ and $\zeta$ represents shear and bulk viscosity of SIDM. The shear stress satisfies the conditions,
 $u_{\mu} \Pi^{\mu \nu} = 0$ and $\Pi_{\mu}^{\mu} = 0$.
 
The
covariant energy momentum
conservation is
\begin{equation}
\nabla_{\mu} T^{\mu \nu} = 0.
\end{equation}
Considering the scalar perturbations in the metric and solving for
   average energy density
   under small fluid velocity approximation, i.e. $\bar{v}^{2}\ll 1$, we get the energy density evolution as
   \cite{Floerchinger:2014jsa} 
\begin{subequations}
	\begin{equation}
	\label{eq:encons}
	\frac{1}{a}\dot{\langle\epsilon\rangle}_{s}+3H\left[\langle\epsilon\rangle_{s}+\left<P\right>_{s}
	-3\langle\zeta\rangle_{s}H\right]  = D,
	\end{equation}
	\begin{equation}
	\label{eq:diss}
	\mathrm{where, }~D = \frac{1}{a^2}\left<\eta\left[\partial_{i}v_{j}
	\partial_{i}v_{j}+\partial_{i}v_{j}\partial_{j}v_{i} -
	\frac{2}{3}\partial_{i}v_{i}\partial_{j}v_{j}  \right]\right>_{s}
	+ \frac{1}{a^{2}}\left<\zeta[\vec{\nabla}\cdot\vec{v}]^2\right>_{s} +
	\frac{1}{a}\left<{\vec{v}}\cdot\vec{\nabla}(P-6\zeta H)\right>_{s}~~.
	\end{equation}
\end{subequations}
In the above expression
$\langle A\rangle_{s}$ represent the spatial average of $A$.
 Here we see that
the evolution of the average energy density ($ \langle \epsilon\rangle_{s} $)
crucially depends on the dissipation term
$D$.
In addition, we need Einstein's equation to
get the equation for the Hubble expansion rate. For this purpose we use 
the spatial average of the trace
 of Einstein equation $\langle G^{\mu}_{\mu}\rangle_{s}
  = - 8\pi G \langle T^{\mu}_{\mu}\rangle_{}$. For
equation of state (EoS) we define
$\langle P\rangle_{s} +\langle\Pi_{B}\rangle_{s} =
\hat{w}_{\mathrm{eff}}\langle\epsilon\rangle_{s}$, and hence we get \textcolor{blue}{
  \cite{Floerchinger:2014jsa}}
 \begin{equation}
 \label{eq:trace}
\frac{\dot{H}}{a}+2H^{2} = \frac{4\pi G \langle\epsilon\rangle_{s}}{3}
 \bigg(1-3\hat{w}_{\mathrm{eff}}\bigg).
 \end{equation}
 Here we find that the dynamics of the energy density and Hubble rate depends on the $D$ and EoS ($ \hat{w}_{\mathrm{eff}}$).
\section{Estimation of Dissipation and Hubble Parameter}\label{sec:dcal}
In this section, we expand upon the ideas discussed in section
  \ref{sec:first} to get an estimate of the dissipation $D$ and set up the
set of equations we solve to get the evolution of universe.
\subsection{Estimating Dissipation $D$}
\label{sec:diss}
To get the Hubble expansion we need to estimate the extent of
    dissipation due to the viscosity of SIDM. To estimate $D$ we take following
  steps:

$(i)$ Viscous coefficients
  $\eta,\zeta$ depend on the thermal averaged
  quantities and hence they depend on the scale over which the
  thermalization has happened.

$(ii)$ To estimate the length scale over which the average
  needs to be taken for viscosity calculation, we make the estimates for
  mean free path. In dilute gas approximation we have $\eta = \rho v \lambda/3$.
  Using (\ref{eq:visc2}) we get $\lambda \sim 10^{10}(1/\rho)(m/\sigma)$ kpc,
  where $\sigma/m$ is in cm$^{2}/$g and $\rho$ is in M$_{\odot}$kpc$^{-3}$
  \cite{Atreya:2017pny}. The $\sigma/m$ estimates for the galactic and cluster
  scales are $\sim 2$ cm$^{2}/$g and $0.1$ cm$^{2}/$g respectively
  \cite{Kaplinghat:2015aga}. The densities for the galactic scales are in the
  range	$\sim 10^{6}-10^{8}$ M$_{\odot}$kpc$^{-3}$
  \cite{Weber:2009pt,KuziodeNaray:2007qi}. This gives mean free path $\sim 1$ 
  Mpc. This is much larger than galactic size ($\sim 10$ kpc). For clusters,
  the densities are $\sim 10^{8}$ M$_{\odot}$kpc$^{-3}$
  \cite{Newman:2012nv,Newman:2012nw} and we get $\lambda \sim 1$ Mpc \cite{Atreya:2017pny}. This means that it is reasonable to assume that dark matter has undergone some
  interactions within a cluster size halo. Thus we assume the cluster size to
  be the smallest scales where averaging needs to done for estimating
  viscosity. Below this scale the hydrodynamic description doesn't hold.

$(iii)$ We assume that dark matter has virialized on this
  length scale ($\sim 1$Mpc) much before our range of interest of redshift
  ($0\le z\le 2.5$) and hence we can assume $\eta$ and $\zeta$ to be constant
  during the evolution of the universe in this redshift range. This assumption breaks down as one goes
 to the epoch of structure formation but that
  is beyond the scope of the present work.
  To calculate the viscosity, we assume the typical cluster scale velocity
  $v \sim \frac{10^{-2}}{3}$ and for $\frac{\langle \sigma v \rangle}{m}$ we
  use the constraint  obtained on the cluster scale as discussed in the Ref.
\cite{Kaplinghat:2015aga}. We have incorporated the case where viscosity
  changes with the redshift in our ongoing work \cite{Mishra:2018}.

$(iv)$ We assume that velocity derivatives are prominent and evolve at a
scale $L$. Thus the scale $L$ should be larger than the scale at which we
estimate viscosity. $L$, in principle, can be any scale between the disspative
scale (where viscosity appear $\sim 1$ Mpc) to the super cluster scale
($\sim 100$ Mpc) where cosmic expansion becomes prominent.

$(v)$ We replace the peculiar velocity gradient $\partial v$ 
by its average
spatial value, i.e. $\partial v  \sim \langle \partial v \rangle_{s} $.  Since the velocity gradients evolve during the
cosmic evolution,  we make an ansatz for the
space average peculiar velocity gradient over the smaller redshift
($ 0\le z \le 2.5$)
\begin{equation}
\langle \partial v \rangle_{s}\sim  \frac{v_{0}}{L}\left( 1+z\right)^{-n} \ ,
\end{equation}
where  $n\geq 0$ is the free parameter. At present $ \langle \partial v \rangle_{s}(z=0) \sim  v_{0}/L$. The parameter $v_{0}$ is the value of velocity on the scale larger than the cluster scale, hence we can assume its value of supercluster scale velocity, i.e. $v_{0}\sim 6v$.

In the light of the above assumptions, we may approximate $D$,
given by eq. (\ref{eq:diss}), as 
\begin{equation}
	D = \big( 1+z\big) ^{2}\left(\frac{v_{0}}{L(1+z)^{n}}\right)^2 \left(  \frac{4}{3}{\eta}  +  2\zeta \right). 
	\label{eq:Dassume}
\end{equation}
From the eq. (\ref{eq:Dassume}) we see that $D$ depends on the
scale where derivatives are most
  prominent ($L$), velocity at scale $L$ ($v_{0}$), viscous
coefficients ($\eta, \zeta$) and redshift ($z$). It is clear from  eq.
(\ref{eq:Dassume}) that in $D$, the contribution from the smaller
scale (where inhomogeneities dominate) is large whereas
contribution from the larger scale (where universe is more or
  less homogeneous) is small.
\subsection{Hubble Rate And Deceleration Parameter}
\label{sec:hcal}
We now set up the
equations for the evolution of Hubble expansion rate and deceleration parameter. 
In term of redshift ($z$), the deceleration parameter $q$, is given by
\begin{equation}
q(z) = -1 + (1+z)\frac{H'}{H}
\label{eq:qdef}
\end{equation}
Using the dimensionless parameter $\bar{H} =  H/H_{0}$, where
$H_{0}$ is the value of Hubble parameter at $z=0$, the eq. (\ref{eq:qdef}) can be rewritten as
\begin{equation}
\frac{d\bar{H}}{dz} =  \frac{(q+1)\bar{H}}{(1+z)},
\label{eq:hevz}
\end{equation}
The equation for deceleration parameter  can be obtained by using the equations
(\ref{eq:encons}) and (\ref{eq:trace})
\begin{equation}
	-\frac{dq}{d\ln a} + 2(q-1)\left(q-\frac{(1+3\hat{w}_{\mathrm{eff}})}{2}\right)
	=  \frac{4\pi G D(1 - 3 \hat{w}_{\mathrm{eff}})}{3H^{3}} 
\label{eq:qevol}
\end{equation}
 It was argued in Ref. \cite{Atreya:2017pny} that for
  the SIDM $\hat{w}_{\mathrm{eff}}\sim 0$. Hence the  evolution eq.
    for $q$, in term of the redshift is written as
\begin{equation}
\frac{dq}{dz} +   \frac{(q-1) \left(2q- 1\right)}{(1+z)} = \frac{4\pi G D}{3(1+z) H^{3}} 
\label{eq:qe}
\end{equation}
where $D$ is given by
eq. (\ref{eq:Dassume}). In terms of the dimensionless
	parameter $\bar{H}$, the above equation
can be rewritten as
\begin{equation}
\frac{dq}{dz} +  \frac{(q-1) \left(2q- 1\right)}{(1+z)} =  \beta \left( \frac{1+z }{\bar{H}^{3}}\right).
\label{eq:qev}
\end{equation}
where 
\begin{equation}
\beta =  \frac{4\pi G }{3H^{3}_{0}}\left(  \frac{4}{3}{\eta}  +  2\zeta \right) \left(\frac{v_{0}}{L(1+z)^{n}}\right)^2. 
\label{eq:beta}
\end{equation}
From eqn. (\ref {eq:qev}) it is clear that the evolution of $q$ depends on $\beta$, the new dissipation parameter.  

Eqns. (\ref{eq:hevz}) and  (\ref{eq:qev}) are  coupled differential equation
in $ q(z) $ and $ \bar{H}(z) $ that need to solve numerically.  The initial condition for $\bar{H}(z)$ and  ${q}(z)$ are given by its present value i.e $\bar{H}(z=0) = 1$ and
$q_{0} = -0.60$ (from CMB observations) \cite{Ade:2013zuv}.
\section{Estimation of model parameter using cosmic chronometer data}
\label{sec:chi}
As we have seen in section \ref{sec:hcal}, the solution for $\bar{H}(z)$
and $q(z)$, depends on two free model parameters $n$ and $L$. In this section, we will 
estimate the best fit value of model parameters $n$ and $L$ of viscous SIDM
model using the $\chi^{2}$ minimization.
\begin{figure}[]
	\centering
	\includegraphics[width=0.7\linewidth]{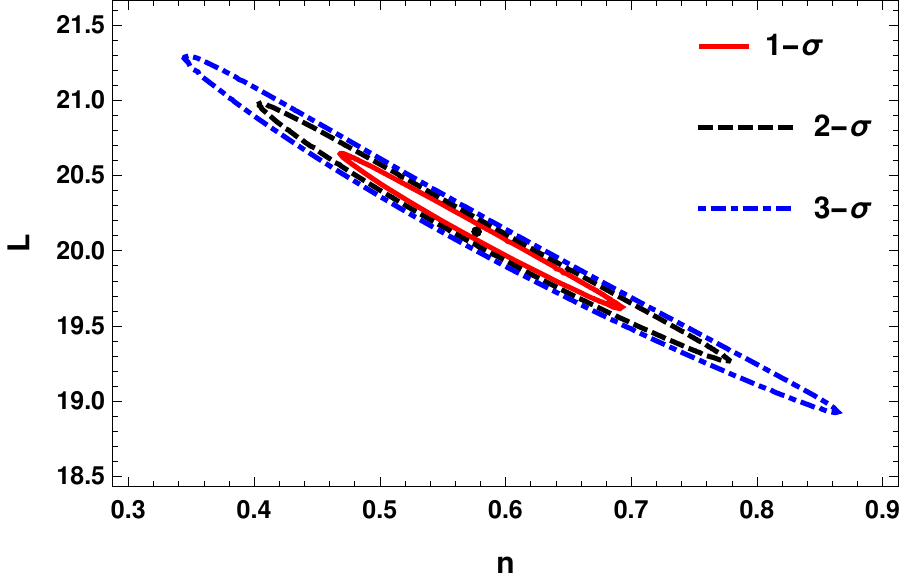}
	\caption{The joint confidence region of model parameters $n$ and $L$ have been plotted. The region correspond to 68.3\%, 95.4\% and 99.73\% confidence limits. The best fit value is shown as a point.}
	\label{fig:cont}
\end{figure}

The theoretical model of the Hubble rate  for
viscous SIDM cosmology is given by the coupled differential eqns.
(\ref{eq:hevz}) and (\ref{eq:qev}). We  find that Hubble rate depends
on two free parameters i.e. $H(z,n,L)$. We use  data for Hubble expansion rate from the cosmic chronometer data set
given in Ref. \cite{Farooq:2016zwm} and reference therein.

The $\chi^{2}$ is defined as
\begin{equation}
  \label{eq:chisq}
\chi^{2}(z,n,L) = \sum_{i=1}^{N}\bigg[\frac{ H_{obs}(z_{i})- H_{th}(z_{i},n,L)}{\sigma_{i}}\bigg]^{2},
\end{equation}
where $N$ is the total number of cosmic chronometer data points and
$\sigma_{i}^{2}$ is the variance in the $i$th  data points. Here $H_{obs}(z_{i})$
and $ H_{th}(z_{i},n,L)\equiv H_{0}\bar{H}_{th}(z_{i},n,L)$ represents  $i$th
observational Hubble parameter data and the theoretically predicted value for
Hubble  parameter respectively.
The best fit value of the model parameters $n$ and $L$ have been estimated
using the $\chi^{2}$ minimization.

The $ \chi^{2} $ per degree of freedom, $ \chi^{2}_{\text{d.o.f}} $, is given by
$\frac{\chi_{min}}{N-M}$, where $M$ is the number of parameters in the model.
In our case $N=38$ and $M =2 $. The
best fit value of model parameters, $ \chi^{2}_{\text{d.o.f}} $ and 1-$\sigma$ confidence
region values are given in the table \ref{tab:best}. The
contour plot of joint confidence region of model parameters $n$ and $L$
corresponding to 68.3\%, 95.4\% and 99.73\%  have 
also been plotted in the Fig. \ref{fig:cont}.

\begin{table}
	\caption{The best fit model parameter }
	\begin{tabular}{||c|c|c|c|c||} 
		\hline
Data set & 1-$ \sigma $ & $  \chi^{2}_{min} $ & $ \chi^{2}_{\text{d.o.f}}$ & Best fit model parameters \\
		\hline\hline
		Cosmic chronometer &  $ 0.5770^{+0.0766}_{-0.0679}$  & 22.02 & 0.61 & $ n=0.5770 $\\
		&  $ 20.1265^{+0.0766}_{-0.3393} $  &  & & $L = 20.1265$ \\
		\hline
	\end{tabular}
	\label{tab:best}
\end{table}
The best fit values are $n = 0.5770$ and $L = 20.1265$ Mpc for the power law
exponent and the gradient length scale. We see that $L$ is approximately an
order of magnitude
larger than the cluster size scale ($\sim $ Mpc), which is the smallest scale
for viscosity estimation.

\section{Results}\label{sec:results}
In this section, we show the our results using the best fit model parameters and also compare with the constant dissipation $n=0$ prediction at the same length scale, i.e. $L=20.1265$ Mpc.
\subsection{Hubble expansion rate}
\begin{figure}[]
	\centering
	\includegraphics[width=0.6\linewidth]{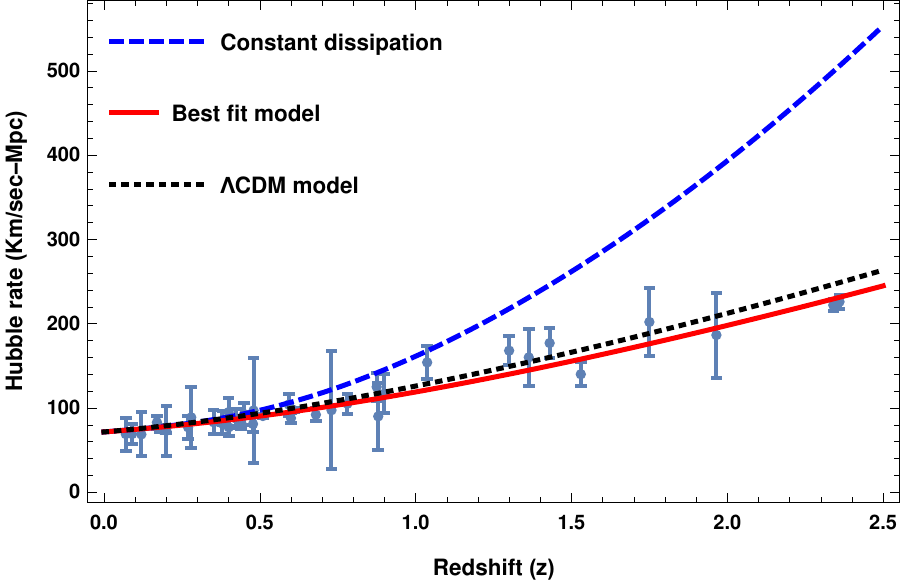}
	\caption{Hubble parameter for the best fit value along with the case of
		constant dissipation is plotted. We also compare it with the
		$\Lambda$CDM model.}
	\label{fig:hbl}
\end{figure}
The plot for Hubble rate, obtained by numerically solving
	eqns. (\ref{eq:hevz}) and (\ref{eq:qev}) and using the best fit
	values of parameters is given in Fig. \ref{fig:hbl}. We also compare it with
	the case of constant dissipation and $\Lambda$CDM model.
	
        We see that constant dissipation case explain the data upto $z\leq 0.7$ and on larger redshift, Hubble start increasing drastically and fails to fit the data. On the other hand, the best fit  and $\Lambda$CDM model explain the Hubble data very well. Thus we find that the constant dissipation case doesn't explain the Hubble data.
\subsection{Fitting of Supernovae data}

Using the Hubble
  rate obtained in section \ref{sec:chi}, we calculate
the luminosity distance, $d_{L}$, given as
\begin{equation}
d_{L}(z) =\frac{ (1+z)}{H_{0}}\int\frac{dz}{\bar{H}(z)}.
\end{equation}

The quantity measured in supernova observations is distance
modulus $\mu$, which is related with the  luminosity distance in the following
manner, 
\begin{equation}
 \label{eq:distmod}
\mu(z)\equiv m-M = 5\log_{10}{\left(\frac{\bar{d}_{L}(z)}{Mpc} \right)} + 25,
\end{equation}
where $\bar{d}_{L}(z) \equiv H_{0}d_{L}(z)$ and $m$, $M$
represent apparent and absolute  magnitude of type Ia supernovae.

\begin{figure}[!t]
	\centering
	\includegraphics[width=0.6\linewidth]{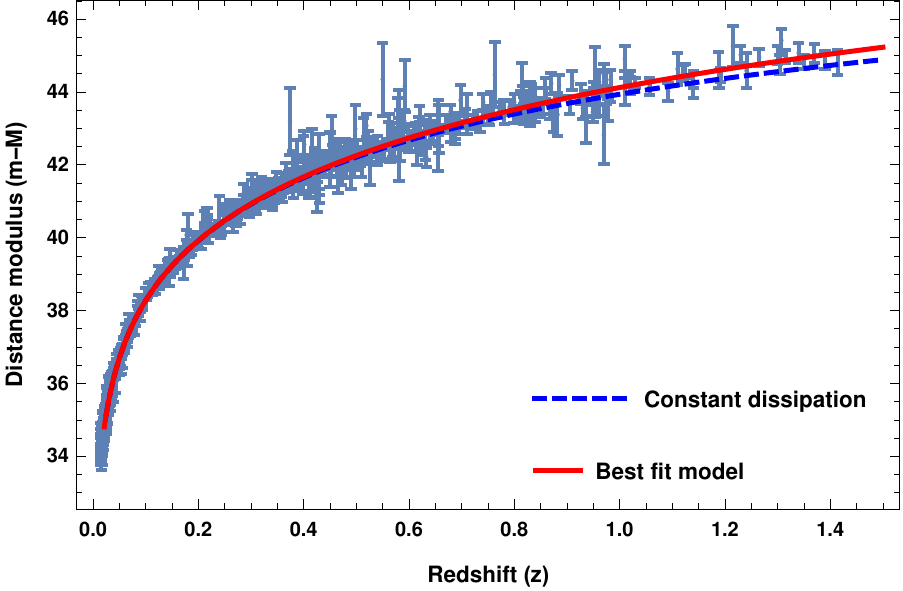}
	\caption{ Distance modulus ($m-M$) obtained from best fit model
          parameters and constant dissipation have been plotted with the
          supernovae data. Our viscous SIDM explains the supernova data}
	\label{fig:dm}
\end{figure}

In Fig. \ref{fig:dm}, we plot the distance modulus ($m-M$) for the best fit
value of model parameters ($n,L$) and plot along with the Supernovae data taken
from the Ref. \cite{Amanullah:2010vv, Suzuki:2011hu}. For a comparison, we
 also plot the distance modulus obtained 
for case of constant dissipation (for $n=0$).
We see that the  best fit and the
constant dissipation case (where the velocity
gradient is constant),  match quite
well on small redshift ($z\le1$), but at large redshift ($z\ge1$),
 the difference between
  the two cases becomes prominent. 
 We thus conclude that the decreasing dissipation with the
redshift can explain both the
cosmic chronometer and the supernova data
while constant dissipations fails to do so.
\subsection{Deceleration parameter (q)} 
\begin{figure}
	\centering
	\includegraphics[width=0.6\linewidth]{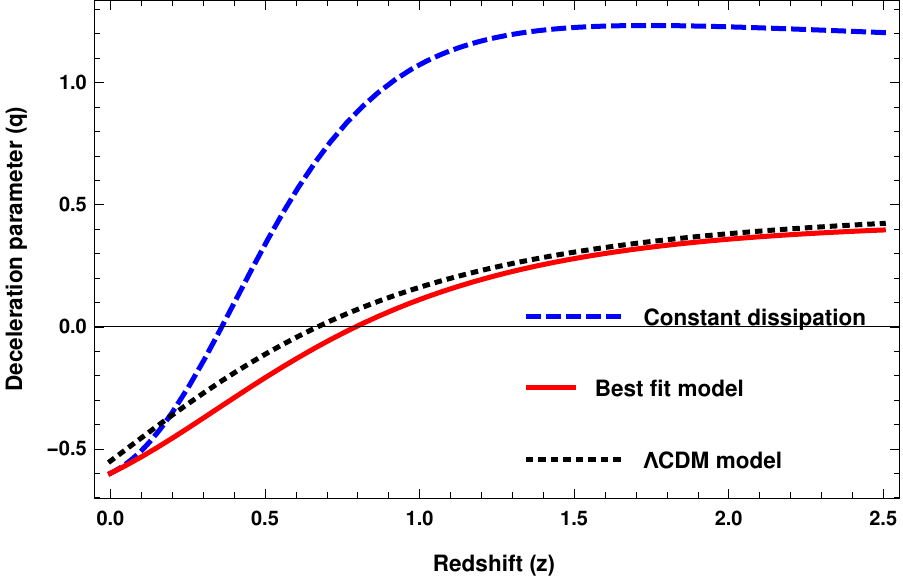}
        \caption{ Plot of deceleration
 parameter ($q$) obtained from best fit  along with the
     $q$ values for $\Lambda$CDM and
constant dissipation case. The best
fit model and $\Lambda$CDM  appraoch $q=0.5$ in the
matter dominated era.}
	\label{fig:q}
\end{figure}

To see the epoch of decelerated to accelerated phase transition (i.e. epoch of
transition from $q>0$ to $q<0$) of the Universe in viscous
SIDM model, we plot $q(z)$ obtained from best fit value of the model
parameters as shown in Fig. \ref{fig:q}. 
To compare our results with the standard $\Lambda$CDM and constant dissipation scenario, we have also plotted the $q(z)$ obtained from the same.

We see that in viscous SIDM model, the transition point $z_{\mathrm{tr}}\sim 0.8$
and in $\Lambda$CDM transition point $z_{\mathrm{tr}} \sim 0.7$.
The transition point of constant dissipation case is
$z_{\mathrm{tr}} \sim 0.3-0.4$ which is later in comparison with the
$\Lambda$CDM and  viscous SIDM model.
We also note that for constant dissipation, the
deceleration parameter increases drastically and settles around $q\sim 1.2$
for higher $z$. This is quite in contrast with our expectation that $q$ should
approach $0.5$ in the matter dominated era.
On the other hand, the value of deceleration parameter $q$ for best fit model
parameter in viscous SIDM model saturates at $ q \sim 0.4 $ which is
approximately same as the $\Lambda$CDM prediction. We may
  thus safely conclude that the case for $n=0$, or constant dissipation, is
  surely not the case to appropriately describe the cosmic evolution.
\subsection{Dissipation parameter ($\beta$)} 
\begin{figure}{H}
	\centering
	\includegraphics[width=0.6\linewidth]{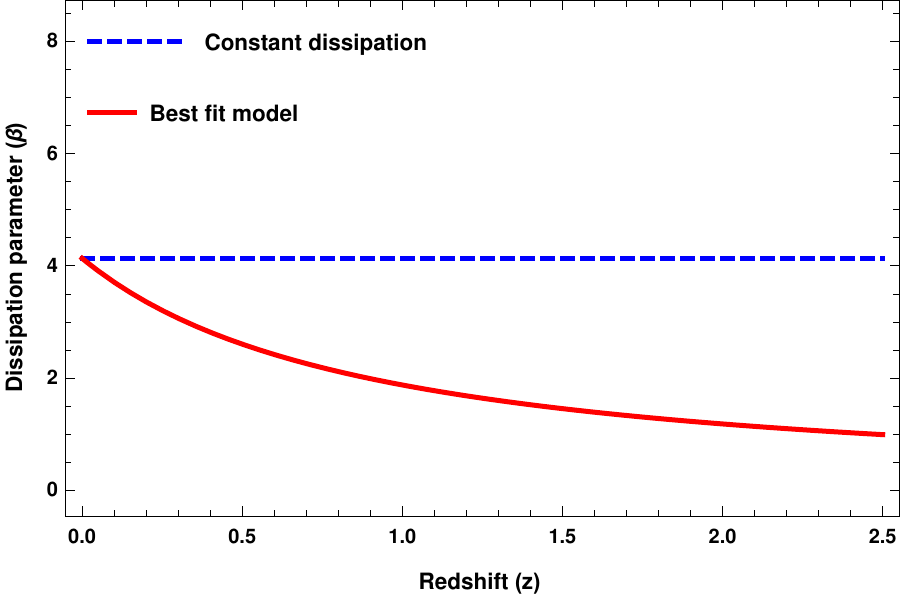}
\caption{ Dissipation parameter ($\beta$) obtained from best fit model parameters of our viscous SIDM model have been plotted along with the constant dissipation model. In farmer the dissipation is decreasing and in later the dissipation is constant with the redshift. }
	\label{fig:diss}
\end{figure}

As we have seen earlier in section \ref{sec:hcal} that the dissipation
parameter $\beta$, contains the information about the dissipative effect of
viscous SIDM due to dark matter self interactions. 
To see the variation of  dissipative effects, we plot the dissipation
parameter ($\beta$) as a function of redshift which is shown in Fig.
\ref{fig:diss} . 

We see that for $n=0$ case, the velocity gradient and viscous coefficients are
constant and hence the $\beta$ is constant. 
Thus the dissipative effects remain prominent even
on the larger redshift. 
On the other hand,
for the best fit value of model parameters that
explain the data well, $\beta$ decreases  sharply and
becomes small
for large $z$. This is not a surprising behaviour
because in past when structures were being formed, the average peculiar velocity
gradients were small,
and consequently the dissipation term was not much effective.
However in the later time of cosmic evolution,
the density perturbations
start growing which in turn
increases the average peculiar velocity  gradient and make the dissipation term important.
Hence we find that the contribution of viscous effects becomes important at a late time not in early time.

From eq. (\ref{eq:qe}) and
  (\ref{eq:beta}) we note that at present epoch $\beta = \frac{4\pi G D}{3H^{3}_{0}}$,
which is $\sim 4.1$ for our case. We see that our present
estimated value, differs 
from the value
($\frac{4\pi G D}{3H^{3}_{0}}\sim 3.5$) reported in 
Ref. \cite{Floerchinger:2014jsa}. Their conclusion was based on the assumption that at the
  present epoch  $\frac{dq}{dz}<<1$.
In our work, by using the simplified assumptions as discussed in section
\ref{sec:diss}, we find
$\frac{dq}{dz} = 0.61$. Hence by incorporating this contribution, we can
explain the discrepancy between the $\beta$ values.

\section{Summary and Conclusion}\label{sec:danc}
The viscous effect due to dark matter self-interactions may play an important
role in the evolution history of the Universe.
In late time the viscous effects of SIDM becomes important and may explain the
present observed accelerated expansion of the Universe and hence mimic
dark energy. In this work, we have tried to explain the cosmic evolution on the redshift
range $0\leq z\leq 2.5$ by incorporating the viscous effects
  in cosmic fluid which arise due to dark matter self interactions.

  Assuming the cluster scale to be virialized and viscous coefficients are
  constant over the  redshift of our interest ($0\leq z\leq 2.5$), we
  calculate the viscosity of SIDM using the Astrophysical constraints.
  Spatial-averages of velocity gradients are modelled to have
    a power law dependence on redshift. With these approximations we
    set up the equations governing the Hubble expansion rate and
  deceleration parameter. We estimate the best fit model parameters ($n,L$) of
  our viscous SIDM  model using $\chi^{2}$ minimization with the cosmic
  chronometer data.
%
%

  We find the best fit values $n=0.5770$ and $L=20.1265$ explain the supernovae data and also attain the
  appropriate $q$ value in matter dominated era. We have also estimated the
  transition point of decelerated to accelerated 
expansion (i.e. epoch of $q>0$ to $q<0$) and found that
  the epoch of transition $z_{\mathrm{tr}} \sim 0.8$. The expectation that velocity gradients
    and hence dissipation should decrease at higher redshift and becomes
  important at  late times in order to
  explain the present observed cosmic acceleration reflects in our viscous
  SIDM model for best fit parameters. Furthermore, we also compare the best
  fit model with the case of constant dissipation
  and find that the constant dissipation does not explain the
data (cosmic chronometer and supernova) and also does not attain the 
  correct value of deceleration parameter in matter dominated era.
  Our analysis is independent of any specific
  particle physics motivated model of SIDM.
\section{Acknowledgement}
We would like to thank Subhendra Mohanty for useful discussions and comments. AM would also like to thank Richa Arya, Sampurn Anand and Jerin Mohan ND for fruitful discussions and suggestions. AA is financially supported
  by Scientific Education Research Board (SERB), under the National
  Post-Doctoral Fellowship (NPDF) grant number PDF/2017/000641.


{} 
 

\begin{thebibliography}{99}
	
\bibitem{Riess:1998cb} 
  A.~G.~Riess {\it et al.} [Supernova Search Team],
  ``Observational evidence from supernovae for an accelerating universe and a cosmological constant,''
  Astron.\ J.\  {\bf 116}, 1009 (1998)
  [astro-ph/9805201].



\bibitem{Perlmutter:1998np} 
  S.~Perlmutter {\it et al.} [Supernova Cosmology Project Collaboration],
  ``Measurements of Omega and Lambda from 42 high redshift supernovae,''
  Astrophys.\ J.\  {\bf 517}, 565 (1999)
  [astro-ph/9812133].



\bibitem{Eisenstein:2005su} 
  D.~J.~Eisenstein {\it et al.} [SDSS Collaboration],
  ``Detection of the Baryon Acoustic Peak in the Large-Scale Correlation Function of SDSS Luminous Red Galaxies,''
  Astrophys.\ J.\  {\bf 633}, 560 (2005)
  [astro-ph/0501171].



\bibitem{Komatsu:2010fb} 
  E.~Komatsu {\it et al.} [WMAP Collaboration],
  ``Seven-Year Wilkinson Microwave Anisotropy Probe (WMAP) Observations: Cosmological Interpretation,''
  Astrophys.\ J.\ Suppl.\  {\bf 192}, 18 (2011)
  [arXiv:1001.4538 [astro-ph.CO]].



\bibitem{Tegmark:2006az} 
  M.~Tegmark {\it et al.} [SDSS Collaboration],
  ``Cosmological Constraints from the SDSS Luminous Red Galaxies,''
  Phys.\ Rev.\ D {\bf 74}, 123507 (2006)
  [astro-ph/0608632].



\bibitem{Copeland:2006wr} 
  E.~J.~Copeland, M.~Sami and S.~Tsujikawa,
  ``Dynamics of dark energy,''
  Int.\ J.\ Mod.\ Phys.\ D {\bf 15}, 1753 (2006)
  [hep-th/0603057].



\bibitem{Yoo:2012ug} 
  J.~Yoo and Y.~Watanabe,
  ``Theoretical Models of Dark Energy,''
  Int.\ J.\ Mod.\ Phys.\ D {\bf 21}, 1230002 (2012)
  [arXiv:1212.4726 [astro-ph.CO]].



\bibitem{Weinberg:1988cp} 
  S.~Weinberg,
  ``The Cosmological Constant Problem,''
  Rev.\ Mod.\ Phys.\  {\bf 61}, 1 (1989).



\bibitem{Padmanabhan:1987dg} 
  T.~Padmanabhan and S.~M.~Chitre,
  ``Viscous universes,''
  Phys.\ Lett.\ A {\bf 120}, 433 (1987).



\bibitem{Gron:1990ew} 
  O.~Gron,
  ``Viscous inflationary universe models,''
  Astrophys.\ Space Sci.\  {\bf 173}, 191 (1990).



\bibitem{Cheng:1991uu} 
  B.~Cheng,
  ``Bulk viscosity in the early universe,''
  Phys.\ Lett.\ A {\bf 160}, 329 (1991).



\bibitem{Zimdahl:1996ka} 
  W.~Zimdahl,
  ``Bulk viscous cosmology,''
  Phys.\ Rev.\ D {\bf 53}, 5483 (1996)
  [astro-ph/9601189].



\bibitem{Fabris:2005ts} 
  J.~C.~Fabris, S.~V.~B.~Goncalves and R.~de Sa Ribeiro,
  ``Bulk viscosity driving the acceleration of the Universe,''
  Gen.\ Rel.\ Grav.\  {\bf 38}, 495 (2006)
  [astro-ph/0503362].



\bibitem{Avelino:2008ph} 
  A.~Avelino and U.~Nucamendi,
  ``Can a matter-dominated model with constant bulk viscosity drive the accelerated expansion of the universe?,''
  JCAP {\bf 0904}, 006 (2009)
  [arXiv:0811.3253 [gr-qc]].



\bibitem{Mohan:2017poq} 
  N.~D.~J.~Mohan, A.~Sasidharan and T.~K.~Mathew,
  ``Bulk viscous matter and recent acceleration of the universe based on causal viscous theory,''
  Eur.\ Phys.\ J.\ C {\bf 77}, no. 12, 849 (2017)
  [arXiv:1708.02437 [gr-qc]].



\bibitem{Cruz:2018yrr} 
  N.~Cruz, E.~Gonzalez, S.~Lepe and D.~Saez-Chillon Gomez,
  ``Analysing dissipative effects in the $\Lambda$CDM model,''
  arXiv:1807.10729 [gr-qc].



\bibitem{Das:2008mj} 
  S.~Das and N.~Banerjee,
  ``Can neutrino viscosity drive the late time cosmic acceleration?,''
  Int.\ J.\ Theor.\ Phys.\  {\bf 51}, 2771 (2012)
  [arXiv:0806.3666 [gr-qc]].



\bibitem{Gagnon:2011id} 
  J.~S.~Gagnon and J.~Lesgourgues,
  ``Dark goo: Bulk viscosity as an alternative to dark energy,''
  JCAP {\bf 1109}, 026 (2011)
  [arXiv:1107.1503 [astro-ph.CO]].



\bibitem{Mathews:2008hk} 
  G.~J.~Mathews, N.~Q.~Lan and C.~Kolda,
  ``Late Decaying Dark Matter, Bulk Viscosity and the Cosmic Acceleration,''
  Phys.\ Rev.\ D {\bf 78}, 043525 (2008)
  [arXiv:0801.0853 [astro-ph]].



\bibitem{Li:2009mf} 
  B.~Li and J.~D.~Barrow,
  ``Does Bulk Viscosity Create a Viable Unified Dark Matter Model?,''
  Phys.\ Rev.\ D {\bf 79}, 103521 (2009)
  [arXiv:0902.3163 [gr-qc]].



\bibitem{Barbosa:2015ndx} 
  C.~M.~S.~Barbosa, J.~C.~Fabris, O.~F.~Piattella, H.~E.~S.~Velten and W.~Zimdahl,
  ``Viscous Cosmology,''
  arXiv:1512.00921 [astro-ph.CO].



\bibitem{Piattella:2011bs} 
  O.~F.~Piattella, J.~C.~Fabris and W.~Zimdahl,
  ``Bulk viscous cosmology with causal transport theory,''
  JCAP {\bf 1105}, 029 (2011)
  [arXiv:1103.1328 [astro-ph.CO]].



\bibitem{Velten:2011bg} 
  H.~Velten and D.~J.~Schwarz,
  ``Constraints on dissipative unified dark matter,''
  JCAP {\bf 1109}, 016 (2011)
  [arXiv:1107.1143 [astro-ph.CO]].



\bibitem{Anand:2017wsj} 
  S.~Anand, P.~Chaubal, A.~Mazumdar and S.~Mohanty,
  ``Cosmic viscosity as a remedy for tension between PLANCK and LSS data,''
  JCAP {\bf 1711}, no. 11, 005 (2017)
  [arXiv:1708.07030 [astro-ph.CO]].



\bibitem{Anand:2017ktp} 
  S.~Anand, P.~Chaubal, A.~Mazumdar, S.~Mohanty and P.~Parashari,
  JCAP {\bf 1805}, no. 05, 031 (2018)
  doi:10.1088/1475-7516/2018/05/031
  [arXiv:1712.01254 [astro-ph.CO]].



\bibitem{Goswami:2016tsu} 
  G.~Goswami, G.~K.~Chakravarty, S.~Mohanty and A.~R.~Prasanna,
  ``Constraints on cosmological viscosity and self interacting dark matter from gravitational wave observations,''
  Phys.\ Rev.\ D {\bf 95}, no. 10, 103509 (2017)
  [arXiv:1603.02635 [hep-ph]].



\bibitem{Cai:2017buj} 
  R.~G.~Cai, T.~B.~Liu and S.~J.~Wang,
  ``Gravitational wave as probe of superfluid dark matter,''
  Phys.\ Rev.\ D {\bf 97}, no. 2, 023027 (2018)
  [arXiv:1710.02425 [hep-ph]].



\bibitem{Lu:2018smr} 
  B.~Q.~Lu, D.~Huang, Y.~L.~Wu and Y.~F.~Zhou,
  ``Damping of gravitational waves in a viscous Universe and its implication for dark matter self-interactions,''
  arXiv:1803.11397 [astro-ph.HE].



\bibitem{Brevik:2017msy} 
  I.~Brevik, Ø.~Grøn, J.~de Haro, S.~D.~Odintsov and E.~N.~Saridakis,
  ``Viscous Cosmology for Early- and Late-Time Universe,''
  Int.\ J.\ Mod.\ Phys.\ D {\bf 26}, no. 14, 1730024 (2017)
  [arXiv:1706.02543 [gr-qc]].



\bibitem{Floerchinger:2014jsa} 
  S.~Floerchinger, N.~Tetradis and U.~A.~Wiedemann,
  ``Accelerating Cosmological Expansion from Shear and Bulk Viscosity,''
  Phys.\ Rev.\ Lett.\  {\bf 114}, no. 9, 091301 (2015)
  [arXiv:1411.3280 [gr-qc]].



\bibitem{Atreya:2017pny} 
  A.~Atreya, J.~R.~Bhatt and A.~Mishra,
  ``Viscous Self Interacting Dark Matter and Cosmic Acceleration,''
  JCAP {\bf 1802}, no. 02, 024 (2018)
  [arXiv:1709.02163 [astro-ph.CO]].



\bibitem{Spergel:1999mh} 
  D.~N.~Spergel and P.~J.~Steinhardt,
  ``Observational evidence for selfinteracting cold dark matter,''
  Phys.\ Rev.\ Lett.\  {\bf 84}, 3760 (2000)
  [astro-ph/9909386].



\bibitem{Tulin:2017ara} 
  S.~Tulin and H.~B.~Yu,
  ``Dark Matter Self-interactions and Small Scale Structure,''
  Phys.\ Rept.\  {\bf 730}, 1 (2018)
  [arXiv:1705.02358 [hep-ph]].



\bibitem{Kaplinghat:2015aga} 
  M.~Kaplinghat, S.~Tulin and H.~B.~Yu,
  ``Dark Matter Halos as Particle Colliders: Unified Solution to Small-Scale Structure Puzzles from Dwarfs to Clusters,''
  Phys.\ Rev.\ Lett.\  {\bf 116}, no. 4, 041302 (2016)
  [arXiv:1508.03339 [astro-ph.CO]].



\bibitem{Gavin:1985ph} 
  S.~Gavin,
  ``Transport Coefficients In Ultrarelativistic Heavy Ion Collisions,''
  Nucl.\ Phys.\ A {\bf 435}, 826 (1985).

	\bibitem{Kadam:2015xsa} 
	G.~P.~Kadam and H.~Mishra,
	``Dissipative properties of hot and dense hadronic matter in an excluded-volume hadron resonance gas model,''
	Phys.\ Rev.\ C {\bf 92}, no. 3, 035203 (2015)


\bibitem{Weber:2009pt} 
  M.~Weber and W.~de Boer,
  ``Determination of the Local Dark Matter Density in our Galaxy,''
  Astron.\ Astrophys.\  {\bf 509}, A25 (2010)
  [arXiv:0910.4272 [astro-ph.CO]].



\bibitem{KuziodeNaray:2007qi} 
  R.~Kuzio de Naray, S.~S.~McGaugh and W.~J.~G.~de Blok,
  ``Mass Models for Low Surface Brightness Galaxies with High Resolution Optical Velocity Fields,''
  Astrophys.\ J.\  {\bf 676}, 920 (2008)
  [arXiv:0712.0860 [astro-ph]].



\bibitem{Newman:2012nv} 
  A.~B.~Newman, T.~Treu, R.~S.~Ellis, D.~J.~Sand, C.~Nipoti, J.~Richard and E.~Jullo,
  ``The Density Profiles of Massive, Relaxed Galaxy Clusters: I. The Total Density Over 3 Decades in Radius,''
  Astrophys.\ J.\  {\bf 765}, 24 (2013)
  [arXiv:1209.1391 [astro-ph.CO]].



\bibitem{Newman:2012nw} 
  A.~B.~Newman, T.~Treu, R.~S.~Ellis and D.~J.~Sand,
  ``The Density Profiles of Massive, Relaxed Galaxy Clusters: II. Separating Luminous and Dark Matter in Cluster Cores,''
  Astrophys.\ J.\  {\bf 765}, 25 (2013)
  [arXiv:1209.1392 [astro-ph.CO]].



\bibitem{Mishra:2018}
  A.~K.~Mishra,
  Manuscript under preperation.

\bibitem{Ade:2013zuv} 
  P.~A.~R.~Ade {\it et al.} [Planck Collaboration],
  ``Planck 2013 results. XVI. Cosmological parameters,''
  Astron.\ Astrophys.\  {\bf 571}, A16 (2014)
  [arXiv:1303.5076 [astro-ph.CO]].



\bibitem{Farooq:2016zwm} 
  O.~Farooq, F.~R.~Madiyar, S.~Crandall and B.~Ratra,
  ``Hubble Parameter Measurement Constraints on the Redshift of the Deceleration–acceleration Transition, Dynamical Dark Energy, and Space Curvature,''
  Astrophys.\ J.\  {\bf 835}, no. 1, 26 (2017)
  [arXiv:1607.03537 [astro-ph.CO]].



\bibitem{Amanullah:2010vv} 
  R.~Amanullah {\it et al.},
  ``Spectra and Light Curves of Six Type Ia Supernovae at 0.511 < z < 1.12 and the Union2 Compilation,''
  Astrophys.\ J.\  {\bf 716}, 712 (2010)
  [arXiv:1004.1711 [astro-ph.CO]].



\bibitem{Suzuki:2011hu} 
  N.~Suzuki {\it et al.} [Supernova Cosmology Project Collaboration],
  ``The Hubble Space Telescope Cluster Supernova Survey: V. Improving the Dark Energy Constraints Above z>1 and Building an Early-Type-Hosted Supernova Sample,''
  Astrophys.\ J.\  {\bf 746}, 85 (2012)
  [arXiv:1105.3470 [astro-ph.CO]].

 \end{thebibliography}
 \end{document}